# A Method to Discover Digital Collaborative Conversations in Business Collaborations

Antoine Flepp[1,2], Julie Dugdale[2], Fabrice Bourge[1] and Tiphaine Marie-Cardot[1]
[1]*Research Area of Digital Enterprise, Orange Labs, 14000, Caen, France*
[2] *CNRS – LIG, Univ. Grenoble Alpes, F-38000, Grenoble, France*
*{firstname.lastname}@orange.com, Julie.Dugdale@imag.fr*

Abstract: Many companies have a suite of digital tools, such as Enterprise Social Networks, conferencing and document sharing software, and email, to facilitate collaboration among employees. During, or at the end of a collaboration, documents are often produced. People who were not involved in the initial collaboration often have difficulties understanding parts of its content because they are lacking the overall context. We argue there is valuable contextual and collaborative knowledge contained in these tools (content and use) that can be used to understand the document. Our goal is to rebuild the conversations that took place over a messaging service and their links with a digital conferencing tool during document production. The novelty in our approach is to combine several conversation-threading methods to identify interesting links between distinct conversations. Specifically we combine header-field information with social, temporal and semantic proximities. Our findings suggest the messaging service and conferencing tool are used in a complementary way. The primary results confirm that combining different conversation threading approaches is efficient to detect and construct conversation threads from distinct digital conversations concerning the same document.

## 1 INTRODUCTION

Recent years have seen a huge increase in the number of digital applications to promote communication and collaboration. Many companies now have a large suite of digital tools, such as Enterprise Social Networks (ESN), conferencing and document sharing software, as well as standard email and messaging clients; all of which are used to facilitate communication and collaboration among their workers. One of the main outputs of the collaborative activity is a document. The collaborative process leading up to the final version of this document may have entailed in-depth discussions drawing upon the deep and tacit knowledge of the collaborators. However the final document rarely contains any underlying rationale and the rich context in which decisions have been made could be buried in the digital exchanges between collaborators. For a new collaborator, these hidden elements may be the key to quickly understanding the decisions that were made without having to contact the original collaborators. We argue that there is valuable knowledge within these tools, which we refer to as Digital Communication and Collaboration Tools (DCCT), can be used to facilitate future collaboration. We focus on the content and the use of DCCT to extract and rebuild the Collaborative Conversation of Document Production (CCDP). The following section discusses related work on knowledge in business collaborations, the use of digital tools in such a setting, and current works on *conversation threading*: the tree classification of Internet messages. The larger goal of our work is to create a knowledge base that will be used to store valuable reusable knowledge about collaborations; this will make future collaborations more efficient and effective. The work described in this paper concerns enriching the database with meta-data from analysing email exchanges during collaborations and understanding better the complementarity of using email with a conferencing tool. The study focuses on collaborations within the Orange Company. However, we believe that the results will be applicable to other companies or other institutions since the need to enhance collaborations is a common goal and our method of tracing Internet messages can be applied to all situations where email is used. Section 3 presents this proposal and section 4 describes the adopted methodology. The results concerning conversation threads in collaborative business messaging are presented and discussed in Section 5.

## 2 RELATED WORK

### 2.1 Knowledge in Collaborations

*Knowledge Worker* is a concept coined by Peter Drucker to describe a person whose main capital is developing or using knowledge in the workplace (Drucker, 1969). Unfortunately its application to workers is limited by the difficulty in measuring and making knowledge explicit (OECD, 1996). Nevertheless since 1996 the Organisation for Economic Co-operation and Development (OECD) has recommended basing the European economy on *Knowledge* which they consider as a driver of productivity and economic growth, thus showing its importance for businesses and workers.

Tacit knowledge has been described by the idea that "we can know more than we can tell" (Polanyi, 1966). Nowadays knowledge is considered as a dynamic concept, moving along a spectrum from explicitness to tacitness (Flepp et al., 2017).

Although a document's content may be considered as explicit knowledge, the process of understanding the document may be considered as uncovering tacit knowledge (Flepp et al., 2017). We hypothesize that this understanding could be greatly aided if we can capture valuable and reusable information embedded in DCCT that were used by the original CCDP. The evaluation of the relevance and a faster exploitation of extracted information will be the foundation of our knowledge base.

### 2.2 Digital Communication and Collaboration Tools (DCCT)

With the advent of new DCCT many researchers believed that older tools, such as email, would be replaced by newer tools such as Enterprise Social Network (ESN). However this has not been the case (Leclercq-Vandelannoitte et al., 2017) and a superposition effect, called the "Napoleon effect", has been observed (Isaac et al., 2008). This effect refers to stacking digital tools in an organisation without replacing or substituting the old digital tools. ESNs are not replacing the messaging service but are used in a complementary way. For example, (Alimam et al., 2016) showed that there was no correlation between an email graph representing 37 workers' professional relations and an ESN graph of the same workers. Indeed ESNs can be used "to expand a worker's scope of relations for future collaboration or communication". Although it is difficult for a knowledge worker to use an ESN to generate new collaborations (Pralong, 2017), engagement and networking can increase by 15% per year (Lecko, 2018). At Orange, the ESN has been available to approximately 150,000, employees for over two years (Orange, 2015) with 41% active collaborators at the end of 2017. The Orange ESN, called Plazza, allows employees to create a group around business or extra-professional issues. Plazza has the typical features of a social network (comment, "like", messaging), but it also has features to facilitate daily work (project management tool, co-editing of documents, enriched directory).

A frequently used DCCT at Orange is CoopNet, a conferencing tool that is used by 58% of collaborators. An organizer can create Virtual meeting rooms and participants can talk via a conference call and view shared documents or applications. CoopNet has an average of 4,800 teleconferences per day (around 1,200,000 in 2016). This shows the complementarity between different DCCT, e.g. by using email to send an invite or report of a teleconference. (Flepp et al., 2017) also showed that collaborations could depend on the type of the collaboration, the collaborators involved, and the used media.

We believe that each service should not be considered individually in the CCDP, but as one in a suite of DCCT available to the knowledge worker. To extract knowledge of a CCDP we focus on email and its link with a CoopNet; two DCCTs most frequently used at Orange.

### 2.3 Conversation Threading

Conversation threading is a feature provided by DCCT to easily group Internet messages and their replies. Linked Internet messages constitute a conversation thread. However there exist different approaches for conversation threading.

First, messages sent via the Internet adhere to an Internet Message Format that is defined by a set of RFC specifications. RFC 5322 (Resnick, 2008) specifies the syntax for text messages sent via email. The header text in emails provides a way for conversation threading. The header fields "References:" and "In-Reply-To" can be used to identify a conversation thread i.e. storing IDs to retrieve the origin of an email conversation, its replies, forwarding, or CCs. Email servers often use these two header fields to build their own conversation threads, called a Thread-Index (Microsoft, 2013). However a Thread-Index is not always available for email clients (Yeh and Harnly, 2006; Palus and Kazienko, 2010). For example if a new email is sent there will be no associated Thread-Index until the email has been answered, because the Thread-Index is generated by the email server. Therefore, some properties of an Internet message are only available for email clients and not for the server messaging service. To thread conversations from a messaging service it is necessary to use other properties than just the "References:" header field. Therefore, the amount of research on conversation threading, not just on header fields, is increasing (Domeniconi et al. 2016). To thread conversations from a DCCT, three other properties in the email are often used: textual content, collaborators, and the date and time of sending an email (time proximity).

(Palus and Kazienko, 2010) use only two properties (collaborators, and date and time) for email conversation threading. They compare their results with those of (Klimt and Yang, 2004) that also used two properties (textual content and, collaborators). Palus and Kazienko failed to rebuild as many conversation threads as Klimt and Yang. However Palus and Kazienko's results differ by only 2% to 4% depending on the time window. This supports the idea of using properties other than the "References:" header field, such as time proximity.

(Yeh and Harnly, 2006) compare two methods for conversation threading. The first, a header-based method, uses the Thread-Index generated by the messaging service server. The second is a similarity-matching method that combines the three properties noted above: textual content, collaborators and time proximity. The header-based method is easier to implement and it performs reasonably well in most cases. However, the method fails when the Thread-Index is encoded by different server messaging services or when there is no header information. Thus, Yeh and Harnly suggest two advancements. The first is the combination of their two methods described above. The second is the extension of their similarity-matching method with any natural language processing (NLP) tool, because their similarity-matching method does not compare the meaning of a sequence of words but only identical words.

Another work (Domeniconi et al., 2016) is closer to ours. Their approach identifies conversation threads from a heterogeneous pool of internet messages from different DCCTs, such as social networks, emails, forums, etc. Although we did not implement deep learning, as did Domeniconi et al., their work shows the complementary use of different DCCTs. The complementarity is detected on properties of textual content, collaborators and time proximity from different DCCTs. Finally, they suggested a variation of their approach, which is: "the reconstruction of conversational trees, where the issue is to identify the reply structure of the conversations inside a thread".

The innovation we propose is to combine these two approaches (header-field information and social, temporal and semantic aspects) not only to reconstruct conversation threads, but also to identify logical links between distinct conversations.

## 3 PROPOSITION

First, we define a Collaborative Conversation of Document Production (CCDP). We then present a use case of conversation threading at Orange from two DCCTs. We introduce the concept of a Collaborative Conversation Thread as being: a conversation thread based on email header information and similarity matching. We suggest that this complementary use of different approaches of conversation threading will lead to a richer knowledge base than if we used only one conversation threading approach with only one DCCT. Our main contribution is conversation threading in a CCDP, which we believe will be useful to detect conversation threads in different conversations from one or several DCCT.

### 3.1 Collaborative Conversation of Document Production (CCDP)

We define a CCDP on two criteria: information exchanges about the production and sharing of the document, and the complementarity between the DCCTs. For the first criterion we retain three properties to identify conversation threads: a group of collaborators, working on the same topic over a limited time period. For the second criterion, we focus on the complementarity between the two most commonly used DCCTs at Orange: messaging service and its link with a conferencing tool.

## 3.2 Existing Conversation Thread (ECT)

At Orange, the conversation threads of the messaging service and the conferencing tool are independent; consequently they have no common Thread-Index. However, in practice a knowledge worker will often use their messaging service to send invites to a teleconference or to report on a teleconference. Therefore to identify the complementarity of the messaging service and a conferencing tool, we distinguish three types of messages of a knowledge worker messaging service: emails, meetings (as they appear in the messaging service calendar) and meeting notifications (e.g. invites, acceptances). Note that we only focus on meetings containing a link to a teleconference. The final goal is to rebuild ECTs of these three types of messages and find new relevant links between them.

## 3.3 Collaborative Conversation Thread

A collaborative conversation thread is a weighted link made by proximities between two messages from different ECTs. A collaborative conversation thread logically links together different ECTs to constitute a CCDP (Figure 1). For example, if the subject field in an ECT of emails is changed, the messaging service will create a new conversation thread. This happens even if there is a minor modification. We presume that we can find proximities between different ECTs by detecting relevant links between two messages from different ECTs.

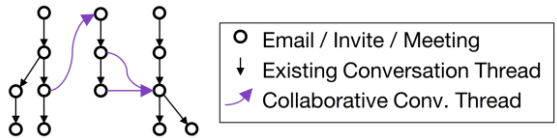

Figure 1: Collaborative conversation thread in a CCDP.

## 4 METHODOLOGY

First we define an equation to link different ECTs. Then we manually apply this equation for a first qualitative evaluation.

## 4.1 Global Proximity (GP)

We calculate a *Global Proximity* (GP) between any two messages of the type *email*, *invite to a meeting*, or *meeting*. If a GP value is above a certain threshold and the two associated messages belong to different ECTs, we consider that we have detected a collaborative conversation thread and we assign these ECTs to the same CCDP.

We define GP as the weighted average of three sub-proximities: *Interlocutors Proximity* (IP), *Time Proximity* (TP), and *Semantic Proximity* (SP), the value of which are all bound in the interval [0,1].

$$GP = (a.IP + b.TP + c.SP) / (a + b + c) \quad (1)$$

Coefficients a, b and c may be used to ajust the impact of any sub-proximity.

### 4.1.1 Interlocutors Proximity (IP)

IP aims to identify how closely the interlocutors are linked through collaboration. A messaging service can identify three different roles of interlocutors in an Internet message. Using the RFC 5322 field names these are: *From*, *To* and *CC*. However we define a fourth role, which is used when an interlocutor is absent from a specific message (*Absent*) while still being part of the conversation. Note that we did not take into account the blind carbon copy role (BCC) because it does not directly concern an active collaboration and it is rarely used at Orange to collaborate. Also, we did not consider the Internet messages sent to oneself since this does not constitute a collaboration. To calculate the IP we attribute an IP coefficient to each combination of roles involved in a conversation (Table 1). Between messages $M_i$ and $M_j$, a coefficient $W_{ij}$ with a value in the interval [0,1] is defined for each pair of messages to reflect how collaborator's roles change.

Table 1: Coefficients of Interlocutors Proximity.

| $M_i$ / $M_j$ | From | To | Cc | Absent |
|---|---|---|---|---|
| From | $W_{ff}$ | $W_{ft}$ | $W_{fc}$ | $W_{f\_}$ |
| To | $W_{tf}$ | $W_{tt}$ | $W_{tc}$ | $W_{t\_}$ |
| Cc | $W_{cf}$ | $W_{ct}$ | $W_{cc}$ | $W_{c\_}$ |
| Absent | $W_{\_f}$ | $W_{\_t}$ | $W_{\_c}$ | |

The combinations involving the two main roles (*From* and *To*) are likely to have higher values than those involving the secondary role *CC*. The combinations involving the *Absent* role have null values (by default) or low values if necessary. After identifying the different interlocutor roles in a conversation and assigning the coefficients we calculate the IP; this is the sum of the IP coefficients

divided by the total number of interlocutors (InterlocNbr) between $M_i$ and $M_j$:

$$IP_{(E_i,E_j)} = \sum W_{xy}(E_i,E_j) / InterlocNbr \quad (2)$$

### 4.1.2 Time Proximity (TP)

We define TP as the difference between the date and time of sending an Internet message ($t_1$) and another one ($t_2$). To bound this proximity in the interval [0,1] we use an exponential inverse function:

$$TP = e^{-(t_2-t_1)/k} \quad (3)$$

Note that k may be modified to have different temporal granularities (e.g. week, day, hour), provided that $t_1$, $t_2$ and k have the same unit.

### 4.1.3 Semantic Proximity (SP)

We use textual semantic proximities between subject-fields, between names of attachments, and between subject-fields and names of attachments. We currently do not take into account email body contents or attachment contents to calculate semantic proximity. Thus we define SP as the result of three semantic sub-proximities: *Subject Semantic Proximity* (SSP), *Attachment Semantic Proximity* (ASP) and *Cross-Semantic Proximity* (CSP).

To calculate these semantic proximities, we use a NLP tool developed at Orange, called SimBow (Charlet and Damnati, 2017). SimBow not only generates a semantic proximity but detects and stores semantic and lexical relations between different bags-of-words. This is useful in our approach where different subject fields in Internet messages may refer to the same topic even if the subject fields have no words in common. For example, "launching" and "closure" are easily recognizable by a human as relating to a project, but are not as easily identifiable to a computer without tools such as SimBow. SimBow supplies a semantical proximity bounded in the interval [0,1]. If two texts are semantically close the calculated proximity tends to 1 (i.e. identical texts have a calculated proximity of 1) and tends to 0 if the two texts are semantically distant. Finally, SP is the maximum proximity found between the three sub-proximities:

$$SP = \max(SSP, ASP, CSP) \quad (4)$$

#### 4.1.3.1 Subject Semantic Proximity (SSP)

To calculate the SSP, we delete the possible prefixes "RE" or "FWD" of subject texts and we use SimBow to give us a semantic proximity between two texts. If at least one of the two messages that are compared has an empty subject text field, the use of SimBow is not necessary and SSP = 0.

#### 4.1.3.2 Attachment Semantic Proximity (ASP)

ASP is calculated in a similar way to SSP. When a message has more than one attachment, we currently concatenate their names into a single string. This simple approach was chosen since it provides us with a first step in semantically comparing attachments. If at least one of the two internet messages to compare does not have any attachment, SimBow needs not be used and ASP = 0.

#### 4.1.3.3 Cross-Semantic Proximity (CSP)

CSP is useful to identify the semantical proximity between a subject text (*subj*) of a message and the name of the attachment (*att*) of another message. For two Internet messages $M_i$ and $M_j$ we define CSP as:

$$CSP_{(M_i,M_j)} = \max[SimBow(subj\ M_i, att\ M_j), \quad (5)$$
$$SimBow(subj\ M_j, att\ M_i)]$$

As with SSP and ASP, if at least one of the two text fields is empty, CSP = 0.

## 4.2 Computation of Proximities

The equations were first tried on a small data set of message exchanges that were selected by a knowledge worker who considered them as a CCDP in his own work.

This CCDP is composed of 3 conversations. A first conversation of 7 exchanges starts with an administrative person requesting the head of a research program to close a project. The latter then transfers the information to the relevant people asking them to prepare a Powerpoint document. The second conversation is composed of two exchanges requesting a teleconference to close the project "officially". The third conversation concerns the invitation to the requested teleconference and the distribution of the meeting report.

Although these three conversations are distinct and have different subjects, they are logically linked as they relate to the same objective ("project closure") and share versions of a Powerpoint document that evolves over time. After the data set was built, four collaborators each subjectively attributed a global proximity measurement to each pair of messages. Note that the collaborators knew about the body content of the messages, whereas our equation does not yet take this into account. We refer to this small sample of subjective global proximities as the Gold Standard. This was calculated by averaging the four subjective global proximities. To evaluate and validate our equation, we compare the Gold Standard to the Calculated Proximities.

## 5 RESULTS

A value of 360 hours for k was found to give the largest standard deviation in time proximities; this best discriminates time proximity values. We then subjectively defined different weighted coefficients to calculate the Interlocutors Proximity (Table 2).

Table 2: Weighted Coefficients for Interlocutors Proximity

| $M_i$ \ $M_j$ | From | To | Cc | Absent |
|---|---|---|---|---|
| From | 1 | 1 | 0.25 | 0 |
| To | 1 | 1 | 0.5 | 0 |
| Cc | 0.25 | 0.5 | 1 | 0 |
| Absent | 0 | 0 | 0 |  |

For the weighted coefficients of the Global Proximity, we currently use an equal distribution (a = b = c = 1/3).

### 5.1 Data Dispersion Analysis

Although the data set is small we analyse the data dispersion in order to compare the difference between the Gold Standard and the Calculated Proximities. We look at the proximities between any two messages globally, between messages that belong to the same conversation thread, or between messages that belong to different conversation threads. Table 3 shows the results at the global level regardless of whether two messages belong or not to the same ECT.

Table 3: Data dispersion between the Gold Standard (GS) and Calculated Proximities (CP).

|  | GS | CP |
|---|---|---|
| Max = | 0.950 | 0.979 |
| Min = | 0.350 | 0.097 |
| Average = | 0.640 | 0.493 |
| Average Absolute Deviation = | N/A | 0.186 |

The average absolute deviation may vary between 0 and 1. Therefore, any value represents an error rate between the Gold Standard and Calculated Proximities, which can be translated in percentage. Hence, with a global level analysis we obtain an average absolute deviation of 0.186, i.e. an 18.6% error rate. Thus, the Calculated Proximities are significantly different from the Gold Standard and so they are not very satisfactory. If we analyse the data further and look at two Internet messages in the same and different ECTs we obtain Table 4 and 5.

Table 4: Data Dispersion between the GS and CP for messages belonging to the same ECT.

|  | GS | CP in a same ECT |
|---|---|---|
| Max = | 0.950 | 0.979 |
| Min = | 0.350 | 0.387 |
| Average = | 0.617 | 0.603 |
| Average Absolute Deviation = | N/A | 0.110 |

Table 5: Data Dispersion between the GS and CP for messages belonging to different ECTs.

|  | GS | CP in different ECTs |
|---|---|---|
| Max = | 0.875 | 0.719 |
| Min = | 0.500 | 0.097 |
| Average = | 0.655 | 0.415 |
| Average Absolute Deviation = | N/A | 0.241 |

The average absolute deviation of 0.186 in Table 3 is largely due to the average absolute deviation between messages of different ECTs, that has a value of 0.241 (Table 5), an 24.1% error rate. Contrarily the fairly low average absolute deviation in Table 3 (11%) indicates that the proximities calculation is efficient between two messages in the same ECT. Nevertheless, the average absolute deviation of 0.241 in Table 5 means that we should improve the calculation of proximities when applying the equation to pairs of messages from different ECTs.

### 5.2 Proximities Balancing

We also analysed the Sub-Proximities. Applying a threshold to the Gold Standard values identified pairs of messages that have the highest subjective proximities. However the data dispersion analysis in Table *3* indicates that we cannot directly compare the Gold Standard with the calculated proximities since the difference between the two sets of data, in particular regarding average values, is too great. Hence we apply a compensatory coefficient to the Gold Standard so that its average equals that of the Calculated Proximities (Table 6).

We should apply the compensatory coefficient to the computed proximities rather than to the Gold Standard (which is the reference). However, some of the CPs would have a value greater than 1, which is meaningless.

Table 6: Data dispersion between the GS and CP after applying a compensatory coefficient.

|  | GS | CP |
|---|---|---|
| Max = | 0.733 | 0.979 |
| Min = | 0.270 | 0.097 |
| Average = | 0.493 | 0.493 |
| Average Absolute Deviation = | 0.124 | N/A |

After applying this compensatory coefficient, the average absolute deviation between the Gold Standard and calculated proximities changes from 0.186 to 0.124, giving a more reasonable comparison between with fewer differences between the dispersions of the datasets.

## 5.3 Proximities Analysis

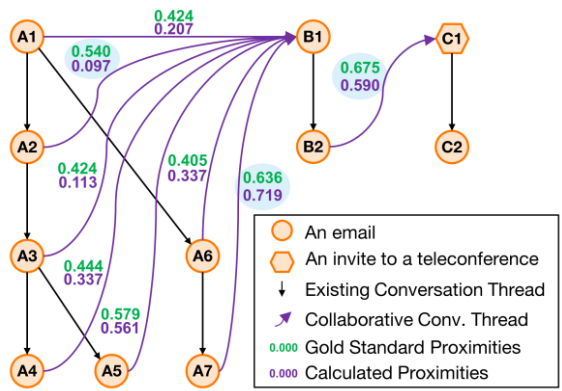

Figure 2: Collaborative Conversation Thread detection.

Figure 2 shows the calculated proximities of the collaborative conversation threads. We focus on the Gold Standard proximities with the highest values between the three conversation threads (A, B and C). An interesting calculated proximity is the one between messages A7 and B1, which belong to different ECTs and share an attachment name identical to the subject of the conversation A. The calculated proximity is 0.719 and the Gold Standard value is 0.636 which is close (less than 0.1).

From the sub-proximities in table 7 the semantic and temporal proximities have a value of 1 and 0.977 respectively, whereas the interlocutors proximity is only 0.179. This means that a collaborative thread has been identified by the calculation of proximities (and verified by the Gold standard) and that this collaboration is due to the messages being semantically and temporally close, and not due to interlocutors' closeness.

Table 7: Example of a temporal and semantic proximities.

| Global Proximity | Interlocutors Proximity | Temporal Proximity | Semantic Proximity |
|---|---|---|---|
| 0.719 | 0.179 | 0.977 | 1.000 |

Moreover, this calculated proximity corresponds to a pair of messages that has the second highest proximity value in the Gold Standard, signifiying that the calculation of proximities is humanly relevant. More precisely, we can see that the calculated proximity between A7 and B1 is greater than calculated proximities between B1 and other messages of conversation A (Figure 2), as knowledge workers done it in the Gold Standard.

A second calculated proximity of interest is the highest proximity value in the Gold Standard between B2 and C1. This calculated proximity is close to the Gold Standard value (0.590 and 0.675, respectively). This is encouraging because it demonstrates that our approach is efficient significantly linking distinct conversations involving collaborations in document production.

Some calculated proximities differ significantly from the Gold Standard as for example with messages A2 and B1 (Table 8):

Table 8: Example of a lack of proximity.

| Messages | Gold Standard | Calculated Proximity |
|---|---|---|
| A2 - B1 | 0.540 | 0.097 |

The calculated proximity differs significantly from the Gold Standard because the equation does not take into account the body content. Therefore it is unable to make any semantic proximity between A2 and B1 (Table 9).

Table 9: Example of a lack of global proximity by analysing the sub-proximities.

| Global Proximity | Interlocutors Proximity | Temporal Proximity | Semantic Proximity |
|---|---|---|---|
| 0.097 | 0.250 | 0.040 | 0.000 |

## 6 DISCUSSIONS & FUTURE WORK

An interesting point is that we are able to identify the most relevant proximities. However, we can alsoidentify less relevant proximities between two messages within the same conversation. This is useful since we will be able to delete those messages that may be irrelevant to collaborations. This helpd to highlight more clearly the elements of the future knowledge base to aid collaboration. A first validation of the equation was done by comparing the calculated proximities with the subjective global proximities. A next step of validation will be to assess if the knowledge base is of practical use to a new member of the collaboration who has not been involved in the final document production. In addition we will extend our approach by: applying the calculation of proximities to more data; using data from different digital communication and collaboration tools, and taking into account the message body content to improve the semantic proximity and thus the global proximity.

## 7 CONCLUSIONS

In this paper, a new approach for conversation threading is introduced that combines a header-field method and a proximity method. Since the header-field method is insufficient for conversation threading on a client messaging service, it is completed by the proximity method using social, temporal and semantic proximities between messages. The results show that our equation can detect that the messaging service and conferencing tool are used in a complementary way during document production. In addition, we confirm that the combination of different approaches of conversation threading is useful to link distinct digital conversations when they concern the same document. These primary results suggest our approach is useful to construct a knowledge base that will aid document understanding.